# Parametric gain and wavelength conversion via third order nonlinear optics a CMOS compatible waveguide

**Alessia Pasquazi, Yongwoo Park, José Azaña, François Légaré, Roberto Morandotti**
*Ultrafast Optical Processing, INRS- Énergie, Matériaux et Télécommunications,*
*1650 Blv. L. Boulet, Varennes, Québec J3X 1S2 Canada*
**Brent E. Little, Sai T. Chu**
*Infinera Corp. 9020 Junction drive Annapolis, Maryland, 94089. USA*
**David J. Moss**
*CUDOS, School of Physics, University of Sydney, New South Wales 2006, Australia email*
*alessia.pasquazi@gmail.com*

**Abstract:** We demonstrate sub-picosecond wavelength conversion in the C-band via four wave mixing in a 45cm long high index doped silica spiral waveguide. We achieve an on/off conversion efficiency (signal to idler) of +16.5dB as well as a parametric gain of +15dB for a peak pump power of 38W over a wavelength range of 100nm. Furthermore, we demonstrated a minimum gain of +5dB over a wavelength range as large as 200nm.

## 1. Introduction

All-optical signal processing is recognized [1, 2] as being fundamental to meet the exponentially growing global bandwidth and energy demands of ultra-high bit rate communication systems. Ultra-fast optical nonlinearities for critical signal processing functions have been widely explored in the last two decades, including four wave mixing (FWM) via the Kerr nonlinearity ($n_2$) for frequency conversion, signal reshaping, amplification and regeneration [3-17].

A key reason for this success has been the development of highly nonlinear waveguides, in silicon [7-12] and nonlinear glasses such as heavy metal oxides [6] and chalcogenide glasses [13-17]. In particular, the first report of gain on a chip was in dispersion engineered silicon nanowires [8], where a net on-chip parametric gain of +1.8dB over 60nm was first reported. Since then, a net on-chip gain of over +30dB was obtained in chalcogenide glass waveguides over a 180nm bandwidth [15]. However, despite this progress, there is still a strong motivation to explore new material platforms in order to achieve the ultimate objective of high nonlinearity together with extremely low linear and nonlinear losses, as well as manufacturability, material reliability and ultimately, CMOS compatibility.

Recently [18-22] we have demonstrated efficient low-power nonlinear optics in very low loss waveguides and ring resonators, in a high index doped silica glass platform. The key advantages of this system are extremely low linear and nonlinear losses together with high reliability and CMOS compatibility. In this paper, we exploit this same platform to demonstrate net parametric gain via degenerate four wave mixing (D-FWM) in a 45cm long spiral waveguide, obtained with sub-picosecond pump and probe pulses. Uniform nonlinear waveguides offer many advantages over resonant structures like ring resonators, such as much wider spectral bandwidths, while at the same time posing different challenges such as requiring larger pump powers - on the order of watts in our case[18], versus milliwatts for ring resonators [19-22].

We achieve a signal to idler conversion efficiency of +16.5dB, as well as parametric gain for the signal of +15dB, with 38W of pump power . While the pumping levels are larger than that for chalcogenide waveguides (our threshold power for 0dB gain is 17W, versus 2W for chalcogenides), our platform has a comparable tuning range (for the same level of gain), and, most importantly, shows no intensity-dependent saturation. Our results are a consequence of extremely low linear (< *0.06dB/cm*) and nonlinear losses (absent up to 25GW/cm$^2$, corresponding to 500W peak power over an effective area of 2μm$^2$ in our device [21]), a high effective waveguide nonlinearity (220W$^{-1}$km$^{-1}$), and near optimum dispersion characteristics (small and anomalous) exhibited by our device in the C-band. The low dispersion also results in a remarkably large bandwidth of almost 200nm (signal to idler), while the high material stability, manufacturability and CMOS compatible fabrication of this integrated platform are attractive features for developing practical devices for systems applications.

## 2. Device

The device under investigation is a 45cm long spiral waveguide with a rectangular cross section core of 1.45 μm x 1.50 μm composed of high index doped silica glass [19-22] ($n_{core}$= 1.7 @ 1550nm) surrounded by silica, on a silicon wafer. The layers were deposited by chemical vapor deposition and the spiral was patterned with high resolution optical lithography followed by reactive ion etching. The 45cm long spiral waveguide is contained within a square area as small as 2.5mm x 2.5mm and it is pigtailed to single mode fibers, with a pigtail coupling loss of 1.5dB/facet. The properties of the material and waveguide dimensions have been engineered to reduce the material dispersion near λ=1550nm, with an anomalous group dispersion ( $\beta_2$ ) over the wavelength range studied here. Measurements of dispersion in this device [21,23] show that for a TE polarization, $\beta_2$ is anomalous for wavelengths shorter than 1600nm, varying from 0 to -20 ps$^2$/km at 1480nm, and so it is ideal (small and anomalous) over most of the L-band, the C-band and indeed well into the S-band. For a TM polarization, the dispersion is also small and anomalous below 1560nm (most of the C-band). From [21,23] we estimate the third order dispersion, $\beta_3$ , to be small and to have the same sign as $\beta_2$ at - 0.3ps$^3$/km. This very wide anomalous dispersion wavelength range enables a very large FWM phase matching tuning range [8,15,24,25].

## 3. Theory

The model used to fit the experimental data is the standard (1+1) nonlinear Schrödinger equation for dissipative-dispersive systems [24]:

$$i\frac{\partial A}{\partial z} - \frac{\beta_2}{2}\frac{\partial^2 A}{\partial T^2} - i\frac{\beta_3}{6}\frac{\partial^3 A}{\partial T^3} + \gamma\ |A|^2 A + i\frac{\alpha}{2} A = 0 \tag{1}$$

where *A(z,t) is* the optical envelope, *z* is the propagation coordinate, *T* is a moving time reference defined as *T=t-z/v$_G$* (here t is the temporal coordinate and *v$_G$* the group velocity). The parameters *β$_2$* and *β$_3$* represent the second and third order dispersion, respectively, while *γ* is the effective nonlinearity and *α* is the attenuation in the spiral waveguide. We note that the losses of the input and output fiber pigtails were also accounted for when solving Eq. (1), which was integrated via a standard pseudo spectral approach. The dispersion coefficients were obtained as a best fit of the group velocity dispersion, reported in [21]. Gaussian pulses were assumed for both the input pump and signal envelopes.

## 4. Experiment

Sub-picosecond pulses for the pump and the signal were obtained from an OPO system (OPAL, Spectra Physics Inc,) generating 180fs long pulses at a repetition rate of 80MHz. The broadband pulse source (bandwidth = 30nm) was split and filtered by two tunable Gaussian bandpass filters operating in transmission, each with a -3dB bandwidth of 5nm, in order to obtain synchronized and coherent pump and signal pulses at two different center wavelengths, each with a pulsewidth of ~700fs. The pump and signal pulses were then combined into a standard SMF using a (90/10)% beam splitter and then coupled into the spiral waveguide. Pulse synchronization was adjusted by means of an optical delay line, while power and polarization were controlled with a polarizer and a λ/2 plate. Both pump and probe polarizations were aligned to the quasi-TE mode of the device.

## 5. Results and Discussion

Fig. 1 shows the measured spectral power densities at the output of the waveguide for a pump wavelength $\lambda_{pump}$=1525nm and three signal wavelengths at $\lambda_{signal}$=1480nm; 1490nm; and 1500nm, respectively. The pump peak power coupled inside the waveguide was varied between 3 and 38W, while the signal peak power was kept constant at 3mW. It is clear that the signal was efficiently converted and amplified into an idler in the C band at wavelengths of $\lambda_{idler}$=1578nm; 1565nm; and 1547nm for all three signal wavelengths.

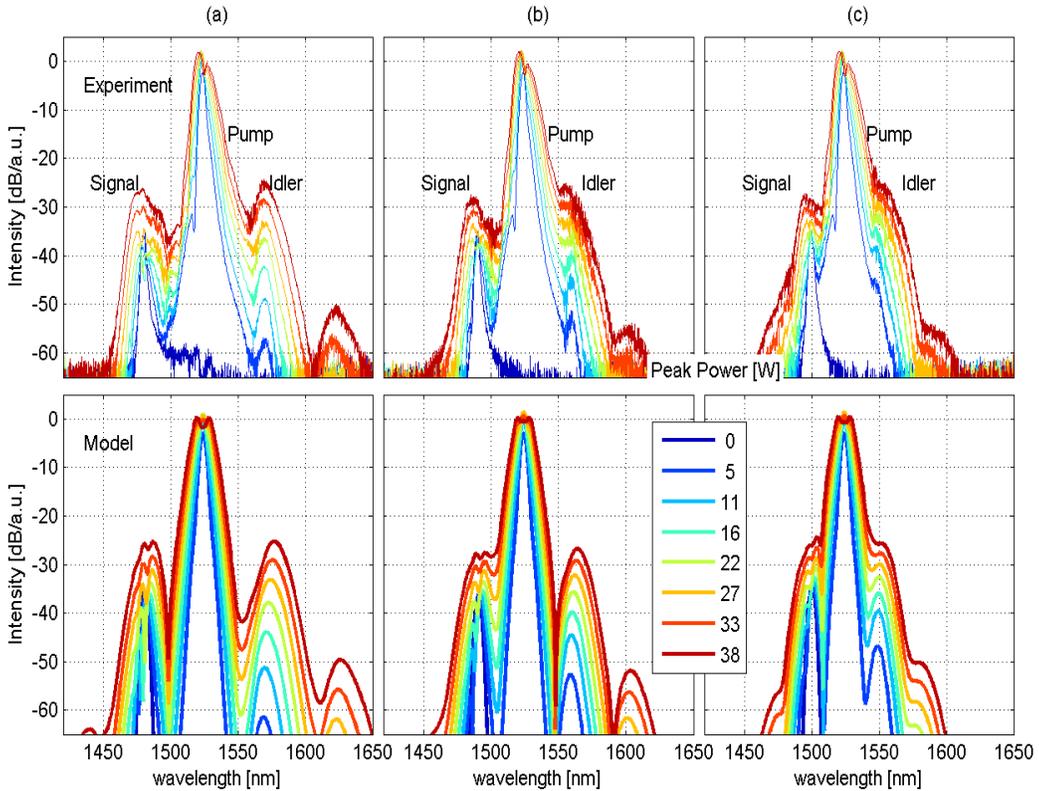

Fig. 1: Experimental (top) and theoretical (bottom) signal intensity spectra for a 1525nm pump and a 1480 nm (a), a 1490 nm (b), and a 1500nm (c) signal. The legend lists pump peak powers (in pseudocolors from blue to red for increasing powers), while the signal peak power was kept to a constant value of 3mW.

The small discrepancy between theory and experiment in the spectra arises from the non-ideal Gaussian pulses used in the experiments. This was due to a number of factors including the deviation from a Gaussian spectral profiles in the low intensity wings of the spectra as well as to residual chirp on the pulses (see below).

For pump powers larger than 25W, despite the fact that we found the idler and signal pulses to be comparable in power, we observed cascaded DFWM only on the idler side at 1625nm. This indicates that the cascaded interaction between the pump and signal pulses at

1525nm -1578nm (producing a cascaded signal at 1625nm) is phase matched, while the other interaction between the pump and the idler at 1425nm and 1480nm (generating light at 1525nm) is not phase matched. Hence, the dispersion of the refractive index relative to the center wavelength (1525nm) must be asymmetric, implying a significant contribution from $\beta_3$, assisted by the low absolute value of $\beta_2$ [21]. As previously addressed, the variation of the experimentally measured GVD dispersion [21] in our device indicates a $\beta_2 < 20\text{ps}^2/\text{km}$ over the wavelength range considered here, with a $\beta_3$ on the order of - 0.3ps$^3$/km (with the same sign as $\beta_2$). Our numerical analysis confirms that this value is consistent with the experiments (Fig. 1).

Fig. 2 compares theory with the experiments for a 1480nm signal and a 1525nm pump. Note that all of the theoretical curves in Fig. 2, except for Fig. 2b), include pulse walk-off effects. Fig. 2a shows the output spectra for a 40W pump, and clearly shows a good agreement with theory based on ideal transform limited Gaussian input pump and signal pulses with a spectral width of 5nm. As was the case for Fig. 1, the slight deviation between theory and experiment is the result of a deviation of the pump spectrum from an ideal Gaussian profile.

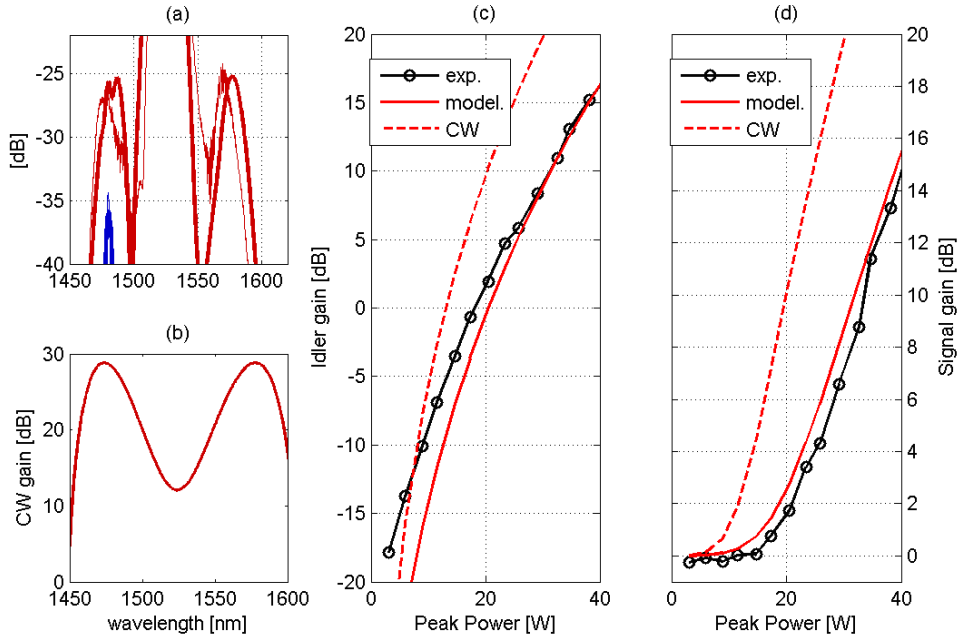

Fig. 2. Gain for a 1480nm signal: (a) Experimental Spectra for a 3mW peak power signal alone (blue) and with a 40W pump (red, thin line), and thick line: numerical model including pulse walk-off effects. (b): Theoretical CW gain. (c-d) FWM gain for idler and signal respectively: measurement (black dots), model in the experimental conditions, pulsed regime (red continuous line) and for a CW regime – red dashed line (note, the CW represents the maximum achievable gain for the pulsed case).

For comparison, we also show the theoretical CW gain spectrum in Fig. 2b, which is noticeably higher than the pulsed case, and which represents an upper limit for the pulsed FWM where walk-off between the pump and signal pulses reduces the conversion efficiency. Figs. 2c and 2d show the peak FWM gain for the idler and signal respectively – ie., the wavelength conversion efficiency (to the idler) and the parametric gain of the signal. The experimental results (black dots) show good agreement with theory (solid red line) that includes walk-off, and both of these are reduced somewhat from the theoretical CW curve (dashed red line) where pump/signal pulse walk-off is absent. Here again, the slight deviation between theory and experiment below pump powers of 25W, observed in Fig. 2(c), is the result of a deviation of the pump spectrum from the ideal Gaussian profile, as well as of a small residual chirp that induces an asymmetric self phase modulation of the pump, visible in both Figs. 1 and 3 (below).

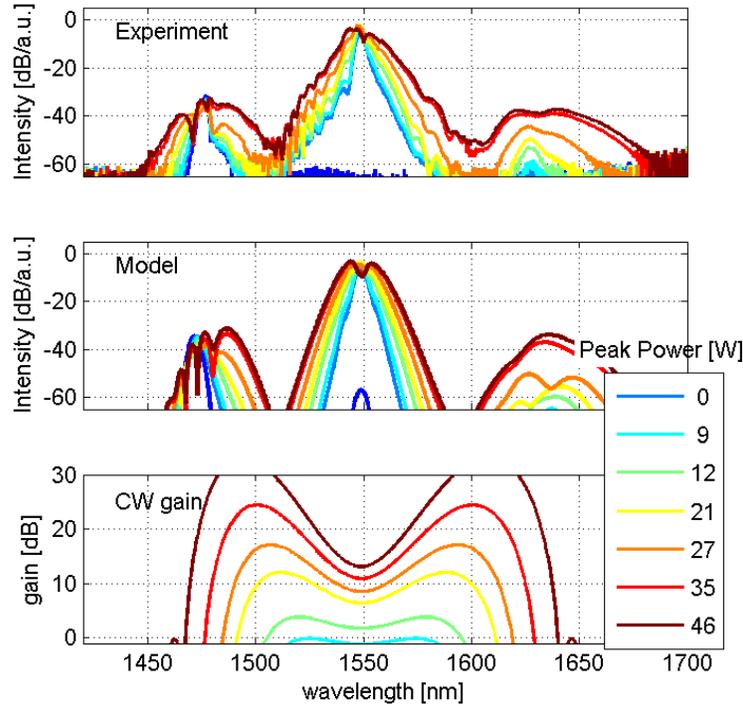

Fig. 3: Ultra-short pulse generation, seeding the FWM interaction at the edge of the CW gain. Top: Experimental output spectrum for a 1550nm pump and a 1475nm signal. Middle: Theoretical output spectrum. Bottom: theoretical CW gain. The legend lists pump peak powers, while the signal peak power is fixed at 15mW.

The experiments (Figs. 1 and 2) show that the FWM bandwidth becomes wider and flattens, as the pump power increases. For pump powers < 22W, there is quite a sharp roll-off in conversion efficiency as a function of the pump-signal separation, whereas for higher powers >30W, the gain becomes much flatter (and larger). This is also observed in the theoretical CW gain (Fig. 2b) for a 38W pump, where the two gain peaks are separated by more than 100nm, with a -10dB bandwidth for each lobe of > 50nm.

We define the "on/off" conversion efficiency as the ratio of the transmitted pulse *energy*, rather than peak power, of the idler (signal) to the transmitted signal without the pump [8]. This allows us to account for the spectral broadening due to XPM, which lowers the spectral intensity. The net, or "on-chip", gain is then the "on/off" gain minus propagation losses. The experimental on/off efficiency and parametric gain vs. pump peak power are shown in Fig.2(c-d), for $\lambda_{signal}$=1480nm, along with theoretical calculations for both a CW and a pulsed pump. For a 38W pump power, we measured a maximum on/off FWM conversion efficiency of +16.5dB from signal to idler, and a parametric gain of the signal of +15dB. This translates into a net on-chip conversion efficiency of +13.7dB and a gain of +12.3dB, when the overall propagation loss of 2.7dB is included. It is important to note that even at the highest pump powers used in these experiments we do not observe any sign of saturation.

Fig. 3 shows the results of experiments with very wide signal to pump wavelength spacing, for a pump wavelength of 1550nm and a signal wavelength of 1475nm, with a signal peak power of 15mW and a pump peak power varied up to a maximum value of 46W. The FWM idler spectrum is larger than 30nm for pump powers above 45W, showing a +5dB gain even when seeded at the edge of the FWM gain spectrum. The large tuning range of the FWM gain process allows for the generation of an idler with a wide spectral width, as well as a clear separation of the idler and pump spectra. Numerical analysis shows that the small 3$^{rd}$ order dispersion term ($\beta_3$) contributes to a red shift of the idler, as well as to the asymmetric nature of the cascaded FWM discussed previously. The interplay between FWM and XPM broadens the spectral bandwidth when the signal is near the edge of the FWM gain curve. While small, this red-shift could be used to help separate the idler and pump, with transform limited pulses (< 100fs) being recovered by subsequently propagating the idler in a negatively dispersive element, as demonstrated recently [26] for SPM.

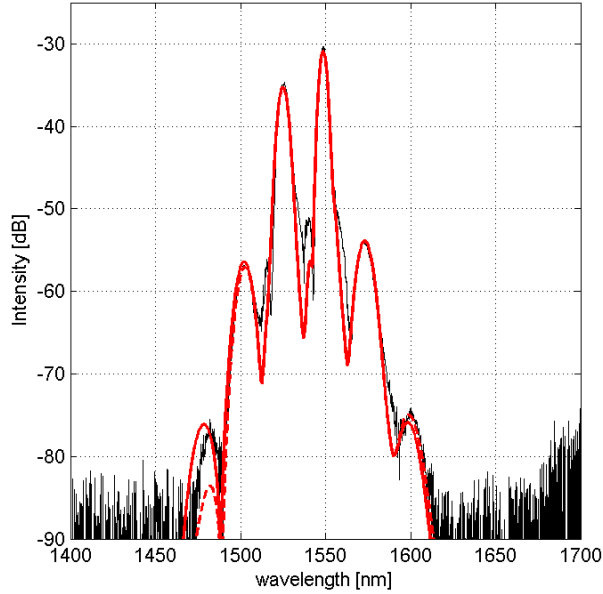

Fig. 4: Cascaded FWM for dual pump pulses at powers and wavelengths of 1.2W at 1525nm and 2.4W at 1550nm, respectively. Black solid line: experiment, Red dashed line: theory without $\beta_5$ and Red solid line: theory with an effective $\beta_5$ of the order of $10^{-2}$ ps$^5$/km.

Finally, we performed experiments with two pump pulses at 1525nm and 1550nm in order to study nonlinear parametric processing in strongly cascaded FWM conditions (Fig. 4). When seeded with two pulses of comparable intensity, FWM can produce a cascade of optical pulses with a well defined relation in frequency and phase [27, 28], which is potentially interesting in wavelength division multiplexing optical communication (WDM) and in optical metrology. The two pump configuration has also been proposed and demonstrated to achieve parametric gain of a weak signal over very wide and flat bandwidths [5,29], including a recent demonstration in a silicon nanowire [30], by pumping near the zero dispersion point. In our case, however, we could not demonstrate this since we did not have access to a separate signal pulse. Nonetheless, with dual pump pulses we managed to achieve cascaded FWM (Fig. 4) generation of secondary and even tertiary idlers with pump powers of only 1.2W and 2.4W at 1525nm and 1550nm, respectively. The theoretical output spectrum (red dashed line) shows a weaker short wavelength secondary idler than what we observed in the experiment (black line), and we found that by including a fifth order dispersion coefficient $\beta_5$ of $+10^{-2}$ ps$^5$/km the agreement between theory and experiment is improved (red solid line) significantly. This method of estimating $\beta_5$ is novel and useful, since this parameter is normally not easy to measure – particularly in integrated waveguides.

## 6. Conclusions

We have demonstrated net parametric gain and a high wavelength conversion efficiency for four-wave mixing in a 45cm high index doped silica glass spiral waveguide. We achieve wavelength conversion over a > 100nm wavelength range using sub-picosecond optical pump and probe pulses, for pump peak powers of ~ tens of watts. We achieve an on/off parametric gain of +15dB and a wavelength conversion efficiency of +16.5dB for a pump peak power of 38W. We explored the generation of large bandwidth optical pulses for ultrafast all-optical applications. Cascaded FWM via dual pump excitation (with peak powers near 2W) was also investigated, and the generation of secondary idlers was achieved. This first demonstration of low power parametric gain via FWM in a CMOS compatible doped silica glass waveguide is promising for all-optical ultrafast signal processing applications, such as frequency conversion, optical regeneration, and ultrafast pulse generation.

## Acknowledgments

We would like to acknowledge support from the Australian Research Council (ARC) including the ARC Centre of Excellence program, as well as from the Natural Sciences and Engineering Research Council of Canada (NSERC). Alessia Pasquazi acknowledges the support of "Le Fonds québécois de la recherche sur la nature et les technologies" (FQRNT), through the "Ministère de l'Éducation, du Loisir et du Sport du Québec" ( MELS) fellowship.

## References and links

1. *Nature Photonics Workshop* on the Future of Optical Communications; Tokyo, Oct. 2007. www.nature.com/nphoton/supplements/techconference2007
2. B. J. Eggleton, S. Radic, and D. J. Moss, "Nonlinear Optics in Communications: From Crippling Impairment to Ultrafast Tools", Chapter 20 (p759-828) in Optical Fiber Telecommunications V: Components and Sub-systems, Edited by I. P. Kaminow, T. Li, and A. E. Willner, Academic Press, Oxford, UK, February (2008).
3. E. Ciaramella, and S. Trillo, "All-optical signal reshaping via four-wave mixing in optical fibers", IEEE Photon. Technol. Lett. **12**, 849-851 (2000).
4. V.G. Ta'eed, L.Fu, M.Rochette, I.C. M. Littler, D.J. Moss, and B.J. Eggleton, "Error-Free All-Optical Wavelength Conversion in Highly Nonlinear As-Se Chalcogenide Glass Fiber", Optics Express **14,** 10371-10376 (2006).
5. S. Radic, C. J. McKinstrie, A. R. Chraplyvy, G. Raybon, J. C. Centanni, C. G. Jorgensen, K. Brar, and C. Headley, "Continuous-wave parametric gain synthesis using nondegenerate pump four-wave mixing," IEEE Photon. Technol. Lett. **14**, 1406-1408 (2002).
6. J. H. Lee, K. Kikuchi, T. Nagashima, T. Hasegawa, S. Ohara, and N. Sugimoto, "All-fiber 80-Gbit/s wavelength converter using 1-m-long Bismuth Oxide-based nonlinear optical fiber with a nonlinearity gamma of 1100 $W^{-1}$ $km^{-1}$ ", Opt. Express **13**, 3144-3149 (2005).
7. H. Fukuda, K. Yamada, T. Shoji, M. Takahashi, T. Tsuchizawa, T. Watanabe, J.-I. Takahashi, and S.-I.Itabashi, "Four-wave mixing in silicon wire waveguides", Opt. Express **13**, 4629-4637 (2005).
8. M. A. Foster, A. C. Turner, J. E. Sharping, B. S. Schmidt, M. Lipson, and A. L. Gaeta, "Broad-band optical parametric gain on a silicon photonic chip", Nature **441**, 960-963 (2006).
9. R. Salem, M. A. Foster, A. C. Turner, D. F. Geraghty, M. Lipson, and A. L. Gaeta, "Signal regeneration using low-power four-wave mixing on silicon chip", Nature Photonics **2**, 35–38 (2007).
10. B. Jalali, D.R. Solli, and S. Gupta, "Silicon's time lens", Nature Photonics **3**, 8-10 (2009).
11. R.L. Espinola, J. I. Dadap ,R. M. Osgood , S.J. McNab and Y.A. Vlasov , "Raman amplification in ultrasmall silicon-on-insulator wire waveguides", Optics Express **12,** 3713-3718 (2004).
12. H.S. Rong ,S.B. Xu, O. Cohen, O. Raday , M. Lee, V. Sih, and M. Paniccia "A cascaded silicon Raman laser", Nature Photonics **2**, 170-174 (2008).
13. V. G. Ta'eed, M. R. E. Lamont, D. J. Moss, B. J. Eggleton, D. Y. Choi, S. Madden, and B. Luther-Davies, "All optical wavelength conversion via cross phase modulation in chalcogenide glass rib waveguides", Opt. Express **14**, 11242-11247 (2006).
14. V. G. Ta'eed, M. D. Pelusi, B. J. Eggleton, D. Y. Choi, S. Madden, D. Bulla, and B. Luther-Davies, "Broadband wavelength conversion at 40 Gb/s using long serpentine $As_2S_3$ planar waveguides", Opt.Express **15**, 15047-15052 (2007).
15. M. R. Lamont, B. Luther-Davies, D-Y Choi, S. Madden, X. Gai, and B. J. Eggleton, "Net-gain from a

parametric amplifier on a chalcogenide optical chip", Opt. Express **16**, 20374-20381 (2008).
16. M. Rochette, L.B. Fu, V.G. Ta'eed, I.C.M. Littler, D.J. Moss, and B.J. Eggleton, "2R Optical Regeneration: Beyond Noise Compression to BER Reduction", IEEE Journal of Selected Topics in Quantum Electronics, Special Issue on All-Optical Signal Processing **12**, 736 (2006).
17. V.G. Ta'eed, M.Shokooh-Saremi, L.B.Fu, D.J. Moss, M.Rochette, I.C.M. Littler, B.J. Eggleton, Y. Ruan and B.Luther-Davies, "Integrated all-optical pulse regeneration in chalcogenide waveguides", Optics Letters **30**, 2900-2902 (2005).
18. D. Duchesne, M. Ferrera, L. Razzari, R. Morandotti, B. E. Little, S. T. Chu, and D. J. Moss, "Efficient self-phase modulation in low loss, high index doped silica glass integrated waveguides", Opt. Express **17**, 1865–1870 (2009).
19. M. Ferrera, L. Razzari, D. Duchesne, R. Morandotti, Z. Yang, M. Liscidini, J. E. Sipe, S. Chu, B. E. Little, and D. J. Moss, "Low-power continuous-wave nonlinear optics in doped silica glass integrated waveguide structures", Nature Photonics **2,** 737–740 (2008).
20. L.Razzari, M.Ferrera, D.Duchesne, M.R.E.Lamont, R.Morandotti, B.E Little, S.Chu and D.J. Moss, "CMOS compatible integrated optical hyper-parametric oscillator", Nature Photonics **4,** 41-45 (2009)**.**
21. M. Ferrera, D. Duchesne, L. Razzari, M. Peccianti, R. Morandotti, P. Cheben, S. Janz, D.-X. Xu, B. E. Little, S. Chu, and D. J. Moss, "Low power four wave mixing in an integrated, micro-ring resonator with Q = 1.2 million", Opt. Express **17**, 14098-14103 (2009)
22. A. Pasquazi, R. Ahmad, M. Rochette, M. Lamont, B. E. Little, S. T. Chu, R. Morandotti, and D. J. Moss, "All-optical wavelength conversion in an integrated ring resonator", Opt. Express **18**, 3858-3863 (2010).
23. D.Duchesne, M.Peccianti, M.R.E.Lamont, M.Ferrera, L.Razzari, R.Morandotti, B.E Little, S.Chu and David J. Moss, "Super-continuum generation in 45cm long spiral high index glass waveguide", Optics Express **18**, 923-930 (2010).
24. B.E. Little "A VLSI photonics platform", Opt. Fiber Commun. **2**, 444–445 (2003).
25. G. P. Agrawal, *Nonlinear Fiber Optics* (Academic Press, San Diego, Ca., (2001)).
26. M.Peccianti, M.Ferrera, D.Duchesne, L.Razzari, R.Morandotti, B.E Little, S.Chu and David J. Moss, "Subpicosecond optical pulse compression via an integrated nonlinear chirper", Optics Express **18**, (8) 7625-7633 (2010).
27. J.M. Boggio, M.S. Chavez, E. Myslivets, J.R. Windmiller, N. Alic, S. Radic, "155-nm Continuous-Wave Two-Pump Parametric Amplification", IEEE Photonics Technology Letters **21**, 612-614 (2009).
28. C. J. McKinstrie and M. G. Raymer, "Four-wave-mixing cascades near the zero-dispersion frequency", Opt. Express **14**, 9600–9610 (2006).
29. A. Cerqueira S. Jr, J. M. Chavez Boggio, A. A. Rieznik, H. E. Hernandez-Figueroa, H.L. Fragnito, and J.C. Knight, "Highly efficient generation of broadband cascaded four-wave mixing products", Opt. Express **16**, 2816-2828 (2008).
30. Park J. S., Zlatanovic S., Cooper M L, Chavez-Boggio J M, Divliansky I B, Alic, N, Mookherjea, S, Radic, S, "Two-Pump Four-Wave Mixing in Silicon Waveguides", OSA Conference Frontiers in Optics (FiO), Paper FML2, San Jose CA, October (2009)